\documentclass[]{article}
\usepackage[left=0.75in,top=1.0in,bottom=1.0in,right=0.75in,nohead]{geometry}

\author{Stuart B. Heinrich \\ North Carolina State University \\ sbheinri@ncsu.edu}
\title{Public Key Infrastructure based on Authentication of Media Attestments}

\usepackage[numbers,sort]{natbib}
\usepackage[pdftex]{xcolor,graphicx}
\usepackage{amssymb}
\usepackage[cmex10]{amsmath}
\usepackage{amssymb}
\usepackage{theorem}
\usepackage{verbatim}
\usepackage{cancel}
\usepackage{rotating}
\usepackage{multicol}
\usepackage{boxedminipage}
\usepackage[labelsep=period,font={footnotesize}]{caption} 
\usepackage{subfig}
\usepackage[hyphens]{url}
\usepackage{algorithm}
\usepackage{algorithmic}

\newcommand{\figref}[1]{Fig. \ref{#1}}
\newcommand{\secref}[1]{Section \ref{#1}}

\begin{document}
\maketitle

\begin{abstract}
Many users would prefer the privacy of end-to-end encryption in their online communications if it can be done without significant inconvenience.  However, because existing key distribution methods cannot be fully trusted enough for automatic use, key management has remained a user problem.  We propose a fundamentally new approach to the key distribution problem by empowering end-users with the capacity to independently verify the authenticity of public keys using an additional media attestment.  This permits client software to automatically lookup public keys from a keyserver without trusting the keyserver, because any attempted MITM attacks can be detected by end-users.  Thus, our protocol is designed to enable a new breed of messaging clients with true end-to-end encryption built in, without the hassle of requiring users to manually manage the public keys, that is verifiably secure against MITM attacks, and does not require trusting any third parties.
\end{abstract}

\begin{multicols}{2}

\section{Introduction}

The majority of modern communication systems, including email, chat, text messaging, and video calls, are insecure and offer little guarantee of privacy to users.  According to Google, Gmail users have ``no legitimate expectation of privacy'' \citep{Musil13}.  In general, although most email servers do require SSL encryption, email contents are processed and/or stored in plaintext on the mail servers themselves.

Not only can these email records be subpoenad to court, but recent revelations show that email and chat communications are frequently divulged \emph{en masse} under duress from government surveillance agencies, that the internal communications networks of major providers such as Google and Yahoo have been hacked by NSA surveillance programs, and that in special cases, government agencies may even be strong-arming certificate authorities to comply in subversion of SSL encryption.

There is a growing public awareness and desire for privacy in response to these recent revelations of mass surveillance.  For example, in a recent study of U.S. writers \citep{FDR13}, 85\% said they were worried about government surveillance, 73\% said they had never been as worried about privacy rights as they are today, and 24\% said they have avoided certain topics over email due to fears of NSA spying.

In theory, true privacy can be achieved by using asymmetric public key cryptography: if each user has a public encryption key and an associated private decryption key that nobody else knows, and encryption for a specific recipient is performed by the sender, and decryption is performed by the recipient, then communications providers only relay encrypted messages and will be incapable of reading their contents.  This is known as end-to-end encryption.

End-to-end encryption protocols for email such as PGP \citep{Zimmermann95PGP} and its variations (e.g., GnuPG \citep{GnuPG} and OpenPGP \citep{OpenPGP}) are well known, but these services are not widely adopted.  There are a number of minor technical excuses that might be given, such as difficulty in performing search, overcomplexity, lack of standardization, and difficulty in distributing untampered client software, but the primary reason is almost certainly the lack of adoption due to end-user inconveniences associated with key distribution and management.

With PGP, an individual user must generate a public/private keypair using an external program, and then keep their private key secret but accessible.  Generally the private key is stored as a file on the local computer, which is inconvenient when the user may wish to read their secure email from a different computer.  The second, much more significant problem is key distribution: in order to send an encrypted email to someone, one must first acquire that person's public key.

Although there are a number of public keyservers that offer free registration and lookup of PGP keys associated with an email address (e.g., \citep{PgpServer}), an end user has no way of knowing if the supplied key is legitimate or not.  A keyserver would be capable of supplying incorrect keys in order to faciliate man-in-the-middle (MITM) attacks, which would enable a third party to read, modify, and impersonate users in the future.  Thus, because a keyserver might have ulterior motives, or might be operating under duress from a government agency, they cannot be universally trusted.

An alternative method of key exchange developed for PGP and related methods is the web of trust (WOT) \citep{Zimmermann95PGP}.  Although the WOT is considered a successful and popular means of exchanging public keys among certain circles, the fundamental problem with the WOT is that it requires caution and intelligent supervision by the users.  It is reliant on \emph{all} the users to understand the weaknesses of the protocol, to care about security, and to be mindful and thorough in verifying each others' identities.  These assumptions do not scale well to the general public, and it is fairly easy to subvert the WOT in order to perform targeted MITM attacks (\secref{sec_wot}).  As such, the WOT cannot be universally trusted either.

In light of these well-known weaknesses, perhaps the most common method of obtaining public keys in practice is a manual exchange between users -- for example, by sending the public key over email.  It might seem that this could be automated, and indeed it would be trivial to design a plaintext-based handshake protocol for exchanging public keys, but an observer with the capability to read and modify these initial messages could act as a man-in-the-middle.  Thus, if such a structured protocol were to come into common usage, it is likely that spying agencies would attempt this type of attack, either by forcing email servers into compliance or by hacking into their networks.

In summary, because there are no sufficiently trustworthy methods for automatically obtaining public keys associated with an email address, client applications cannot perform this operation in the background, and hence users must be burdened with the inconvenience of manually exchanging public keys with each new person they wish to communicate securely with.  It is not surprising that the majority of users don't have the patience for this, and thus end-to-end encryption has remained niche.

For end-to-end encryption to gain mainstream traction, a more reliable and convenient method of key management and exchange must be used.  Several companies such as Lavabit and Hushmail have attempted this by generating PGP keys on behalf of their users, thereby allowing users to login with a simple password (rather than requiring a PGP private key), and automatically looking up PGP public keys of in-network recipients using an internal keyserver.  However, the problem with these hybrid approaches is that the service providers have access to users' private keys and can therefore read users' private messages.  This problem has been recently elucidated by the fact that both Lavabit and Hushmail have been forced to comply with court-orders and NSA demands to turn over internal communications, after which Lavabit (but not Hushmail) chose to shut down its services rather than to remain complicit in widespread surveillance of its users.

In response to Lavabit's shutdown, users and secure email services alike have proposed to shift their business outside of the US, to countries with better privacy regulations\citep{Ludwig13}.  We propose a more permanent, and fundamentally different approach, by providing a method for end-users to independently verify the authenticity of a public key by using media attestments.  When end-users have this capability, client software is free to automatically lookup public keys associated with email addresses from a keyserver without needing to trust the keyserver, because any MITM attacks could be detected by end-users.  Thus, our protocol enables a new breed of messaging clients with true end-to-end encryption built in, without the hassle of requiring users to manually manage the public keys for their contacts, and without relying on any third party to keep certain information secret.

We first demonstrate the need for a more secure protocol by explaining the ease in which the WOT may be subverted (\secref{sec_wot}).  Then we introduce the proposed key authentication protocol (\secref{sec_key_auth}) by explaining the inspiration from `ask me anyting' (AMA) sessions on Reddit (\secref{sec_inspiration}) and showing how this theory can be extended for secure key exchange (\secref{sec_overview}).  We detail the proposed protocol steps for key registration (\secref{sec_registration}), lookup (\secref{sec_lookup}), rating (\secref{sec_rating_card}) and removal (\secref{sec_key_removal}).  Finally, we discuss potential attacks and resistances (\secref{sec_attacks}), present our implementation of a public keyserver (\secref{sec_implementation}), propose recommendations for compatible client software (\secref{sec_client_software}), and give closing remarks (\secref{sec_conclusion}).

\section{Web of Trust Limitations} \label{sec_wot}

Public keys are found in the WOT by searching for a chain of signed public key certificates.  Keys which can be found with fewer hops are generally considered to be ``more trustworthy.''  As such, a common metric for trustworthiness of a public key is the mean shortest distance (MSD) to all other keys \citep{Penning13}.  

In the interest of increasing their own MSD, each user has an incentive to sign as many public keys as they can.  This has a detrimental effect on security, because it encourages people to sign public keys for their own apparent benefit, making it significantly easier for an impersonator to get their impostor keys signed.  Moreover, the users who exhibit the least restraint or least thorough verification of identities will be able to sign the most keys, and hence will have the lowest MSD's.  Finally, anyone who is signed by someone with a low MSD will, by proxy, also acquire a low MSD.  Thus, it is very easy for an impersonator to acquire a low MSD, simply by getting their key signed by someone who has a low MSD, which is very likely the same person who is willing to sign a key without doing a thorough background check.

Furthermore, users are encouraged to sign other peoples' key certificates if their identity can be verified without regard to the integrity of that person.  According to \citet{Zimmerman97}, ``You aren't risking your credibility by signing the public key of a sociopath, if you were completely confident that the key really belonged to him.''  Unfortunately, this is simply not true, because a person with ulterior motives can go on to sign keys that they know are false with the effect of making those keys \emph{appear} trustworthy to you or other people based on graph analysis.

For example, if Eve wants to spy on Alice's communications with Bob, she could just get her associates Charlie, Dave and Francis to have their keys signed by Alice and Bob after verifying their identities.  Then those same associates could certify Eve's impostor key as the key for Alice, and also certify another one of Eve's impostor keys for Bob.  After doing this, if Alice tried to look up Bob's key, she would find multiple independent pathways, all with a short link of just 1 hop, leading to the impostor key (and similarly, for Bob trying to find Alice's key).

Despite being so easy to game the system, the WOT has a lot of followers among security experts and privacy enthusiasts that would call it a success based on their own empirical observations that it appears to work.  Of course, it should come as no surprise that the WOT will work to exchange the public keys between two parties, neither of whom are high profile targets for espionage, because hacking the WOT cannot be done \emph{en masse}, and would require some effort be expended for each target.

The general public largely uses unencrypted email, and those with a genuine need for security are likely to exchange keys over more secure channels.  Thus, most targets worth spying on aren't using email encryption, or know better than to use the WOT.  Thus, there is simply little incentive for spying agencies to subvert the WOT.

However, if the WOT were deployed on a massive scale, as a means for exchanging public keys behind all major email services in the background, then this would certainly change.  First, existence of a pathway of introducers on the WOT would become almost meaningless, due to the large number of general users who might be certifying keys with little regard to security.  Second, if an agency wanted to spy on some individual, they would likely undertake means to start spreading impostor keys, possibly using bots to sign key certificates, and they would either succeed in intercepting some user's requests or create sufficient confusion about which key was correct as to effectively serve as a denial of service (DOS) attack against the usage of the WOT.

\section{Key authentication protocol} \label{sec_key_auth}

In this section we introduce the proposed key authentication protocol.  We begin by explaining the inspiration from `ask me anyting' (AMA) sessions on Reddit (\secref{sec_inspiration}), and show how this theory can be applied for secure key exchange (\secref{sec_overview}).  We call this Authentication of Media Attestments (AMA), having an obvious double meaning.  We then detail the proposed protocol steps for key registration (\secref{sec_registration}), lookup (\secref{sec_lookup}), rating (\secref{sec_rating_card}) and removal (\secref{sec_key_removal}).  Finally, we discuss community rating statistics (\secref{sec_cci}) and guidelines for media attestment creation (\secref{sec_guidelines}).

\subsection{Inspiration from Reddit} \label{sec_inspiration}

\begin{figure*}
\centering
\subfloat[]{ \includegraphics[width=0.25 \textwidth]{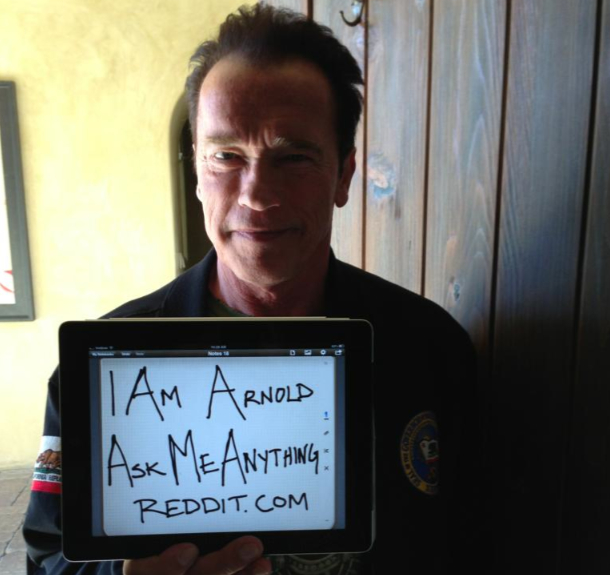}}
\subfloat[]{ \includegraphics[width=0.25 \textwidth]{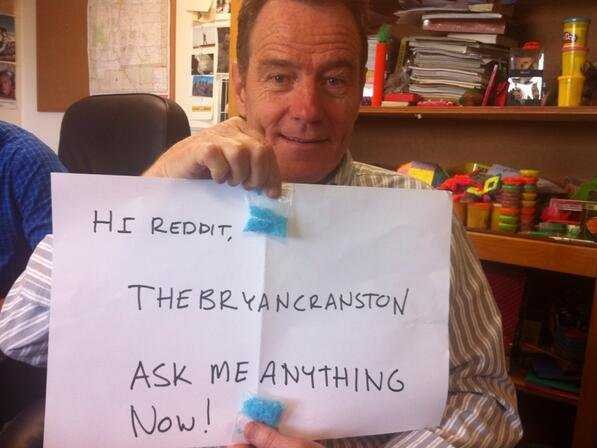}}
\subfloat[]{ \includegraphics[width=0.25 \textwidth]{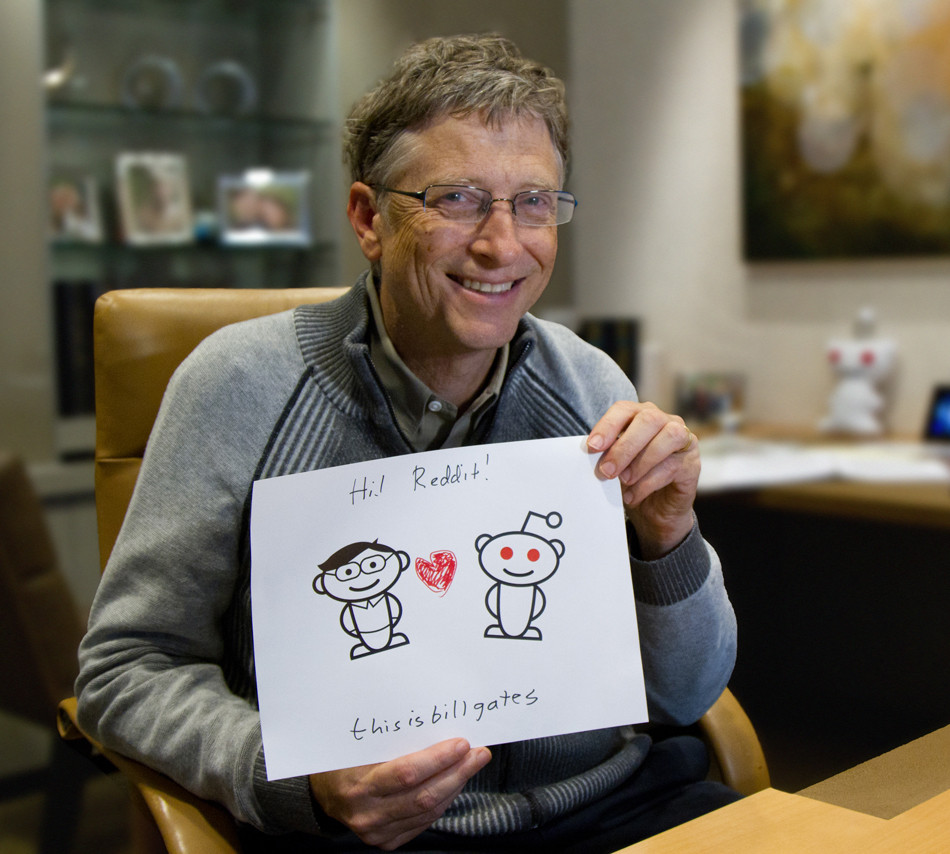}}
\subfloat[]{ \includegraphics[width=0.25 \textwidth]{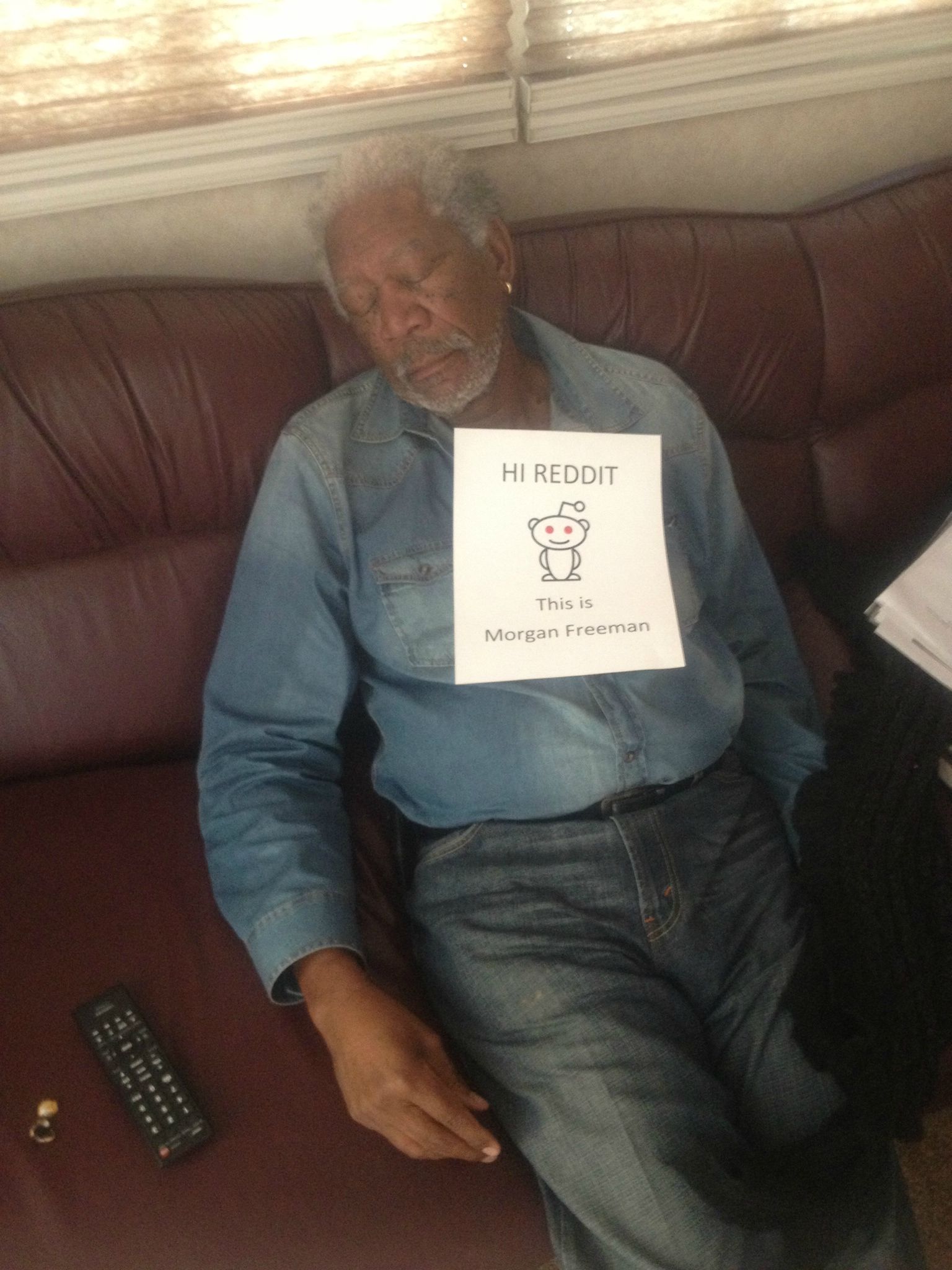}}
\caption{Example of notable AMA's from Reddit. (a) Arnold Schwarzenegger; (b) Bryan Cranston; (c) Bill Gates; (d) Morgan Freeman (fake).}
\label{fig_AMA_examples}
\end{figure*}

Celebrities sometimes make an appearance on the popular geek news site Reddit \citep{Reddit} to answer questions.  In order to authenticate their identities, they sometimes provide a photo of themselves holding a piece of paper indicating that they are doing an AMA session (\figref{fig_AMA_examples}).

These photos are generally convincing because the celebrity is recognizable, and because the piece of paper indicates that an AMA session will be held.  Although the content of a piece of paper could potentially be changed using image editing software, it would generally be difficult to acquire a photo of a celebrity in the posture of holding up a piece of paper that could not be easily recognized as coming from some commercial work in TV or film, which leads one to believe that the photos are authentic.

The fundamental principle behind this type of authentication is that a legitimate photo of oneself is extremely easy to create, and generally easy for others to verify is authentic, but very difficult to forge (although this does not always stop people from trying).  

A notable example is the fake AMA attempted for Morgan Freeman, shown in \figref{fig_AMA_examples}d.  This photo should be immediately suspicious because Freeman is not actually holding the paper, and appears to have simply been photographed while sleeping, and because the text on the paper is not handwritten.  The general consensus (after much analysis on Reddit) is that the piece of paper pictured in the photo was computer generated.

\subsection{Adaptation for secure key exchange} \label{sec_overview}

A similar approach to Reddit AMA-style photos can be used to verify ones public key, with the basic idea being for the true owner to create some media file in which they show some identification and communicate a hash of their public key because a hash is shorter and easier to convey verbally.

It would be insufficient to simply take a photo of oneself holding up a card with the hash written out on it, due to the ease with which a photo might be edited to change the depicted characters in order to perpetrate a MITM attack.  Therefore, we consider video clips, and ask that users communicate their key both visually and verbally, while taking additional measures designed to increase the difficulty of media editing to change the communicated hash value.  We discuss the specific guidelines for creating a video that is difficult to forge in \secref{sec_guidelines}.

A good media attestment effectively proves that an individual is the owner of a public key, and a digital signature from the associated private key can, by proxy, be used to prove that this same individual has authored any other information such as a preferred email address.  Thus, a keyserver can prove that there is no MITM simply by providing a link to the media attestment in addition to the public key.  This is important, and fundamentally different from any current keyservers, because it permits a client to utiliize a keyserver without needing to blindly \emph{trust} the keyserver.

When a client requests the public key associated with a contact address for the very first time, all they need to do is watch the media attestment to see who they will be communicating with.  Because the keyserver must provide the media attestment as proof, and because the media attestment is so difficult to forge, clients may trust that the keyserver is incapable of performing a MITM attack and not wish to bother watching the media attestment at all.

To further facilitate this, we introduce a community rating scheme, whereby users who watch a video attestment can rate the authenticity of this media attestment, sign it with their own private key, and submit this to the keyserver's database.  Then, in the future, the keyserver can supply these signed ratings along with the media attestment itself, so that a client may opt to trust a public key with sufficiently positive community ratings without bothering to actually watch the media attestment.

The proposed protocol is similar to the MITM protection used in the ZRTP protocol \citep{zrtp} for VOIP communications in Silent Circle, although there are some notable differences.  First, because we apply it in a PKI context, a user only needs to read a hash string once (or when they change their password), rather than needing to do it for every voice communication, as in ZRTP.  Second, the proposed media attesments provide MITM protection for any form of communication such as email and text, whereas ZRTP only works for VOIP communication.  Third, the proposed media attestments must follow guidelines involving multiple forms of communication that are much more difficult to forge.  Finally, the persistent nature of our media attestments allow community vetting and rating to bring attention to suspicious attestments.

\subsection{Public key registration} \label{sec_registration}

In this section we describe the protocol for a user to register their public key with a keyserver.

\begin{enumerate}
\item User computes the MD5 hash \citep{Rivest1992} of their public key, which is shorter and more reasonable length to read.
\item The user creates a media attestment of the hash according to the guidelines (\secref{sec_guidelines}).  This is basically a video of them communicating the hash by reading it aloud (as a hex string) and showing or writing out the hash string.
\item For each contact address (ie, email address, mobile phone number, Facebook ID) that the user wishes to associate with their key, they fill out a separate \emph{identity card}.  The identity card contains their public key, a hash of the media attestment (or url, if hosted), and any other optional information such as their name.  Then the user signs the identity card with their private key using a signature scheme such as RSASSA-PSS, and sends all of the signed cards to the keyserver.  Each card only has one contact address so that the server can provide the signed card to a client without revealing the user's other contact methods and addresses.
\item The keyserver then verifies the contact address by generating a nonce (random hex string) and sending it to the supplied contact address as an HTML query parameter to a webserver, which finalizes the registration entry in the database.
\end{enumerate}

\subsection{Public key lookup} \label{sec_lookup}

In this section we describe the protocol for obtaining and validating a public key from a keyserver associated with an email address.

\begin{enumerate}
\item The user sends an email address to query to the keyserver.
\item The keyserver returns the identity card associated with that email address, along with a list of user ratings, and some aggregated statistics based on the user ratings.  For each community rating, the client may lookup the public key for that user from their contact address in order to verify their signature on the rating card, if desired.  Furthermore, the client may verify the aggregated statistics from the collection of individual community ratings, if desired.
\item The client may opt to automatically trust the returned key from on the aggregated statistics and/or content of the individual rating cards, based on some client or user specified thresholds/algorithms.
\item If the conditions for automatic trust are not met, the user may view the included media attestment and ratings, and either confirm or deny the authenticity by filling out and submitting a rating card (as described in \secref{sec_rating_card}).
\end{enumerate}

\subsection{Public key rating} \label{sec_rating_card}

In this section we describe the protocol for one user to submit a rating of another's media attestment:

\begin{enumerate}
\item After viewing the media attestment, the client is presented with the following questions:
\begin{enumerate}
\item Can you recognize the person in this video as the actual owner of the contact address? (yes/no/unsure)
\item Does the hash communicated in the video match the hash given in the identity card? (yes/no/unsure)
\item Does the video meet all mandatory guidelines and appear authentic? (yes/no/unsure)
\item Additional comments? (text)
\end{enumerate}
\item The answers to these questions, along with the rating user's contact address, the original identity card being rated, and the current time, are signed with the client's private key and returned to the keyserver after solving a CAPTCHA \citep{Ahn03captcha} test.
\item The keyserver adds the clients rating to its database and updates the aggregate rating statistics (\secref{sec_cci}).
\end{enumerate}

\subsection{Public key editing/removal} \label{sec_key_removal}

The process for editing a public key registration is simply to remove it and then re-upload a new key.  The process for removing a key record is simple:

\begin{enumerate}
\item User sends a request to remove a contact address signed with their private key.
\item Server verifies that the signature matches the public key on file for that address.  If so, the identity card is deleted.
\end{enumerate}

Alternatively, if the user has lost their private key, they can remove the old public key by verifying ownership of the contact address:

\begin{enumerate}
\item User sends a request to remove the identity card associated with a contact address.
\item Server replies to the contact address with a random nonce.
\item User replies by confirming the nonce, and server deletes the identity card.
\end{enumerate}

\subsection{Aggregated rating statistics} \label{sec_cci}

The following aggregated ratings statistics are computed from individual user ratings, and serve to assist the client software and determining whether or not to trust a public key without bothering to view the media attestment:

\begin{description}
\item[S1] Total number of user ratings.
\item[S2] Count of the number of ratings which confirmed owner's identity.
\item[S3] Count of the number of ratings which denied owner's identity.
\item[S4] Count of the number of ratings which confirmed the correct hash value.
\item[S5] Count of the number of ratings which denied the correct hash value.
\item[S6] Count of the number of ratings which confirmed the video authenticity.
\item[S7] Count of the number of ratings which denied the video authenticity.
\end{description}

It should be noted that these statistics are only provided as a convenience, and can be calculated and verified from the client from the raw user ratings.  Moreover, the client software is free to use any algorithm desired to determine whether or not they want to trust the provided key.  We suggest trusting the key under the conditions that

\begin{align*}
\alpha &< S1  \\
\beta &< \min( S2-S3, S4-S5, S6-S7 ) / S1,
\end{align*}

\noindent where $\alpha \geq 0$ and $\beta \in (0, 1]$ are user-tunable thresholds in the client software.

\subsection{Media attestment guidelines} \label{sec_guidelines}

A media attestment is a video file created by the user in which they communicate a hash of their public key.  It is critical that a media attestment be created in such a way that it would be difficult to modify and change the communicated key, and also difficult for any other person to create.  With these goals in mind, the following guidelines are proposed:

\begin{itemize}
\item The video should be shot in a single take and maintain focus on the users face throughout to prevent the opportunity for an attacker to perform splicing.
\item The user should introduce themself briefly and show some form of ID, such as a drivers license or passport.
\item The user should then speak the hash of their public key aloud, without using a monotone voice, and by reading the digits in short groups so that segmentation of the vocalizations corresponding to individual letters is more difficult.
\item The video should contain some background noise or music that is not easily separable from the vocals, to further increase the difficulty of segmenting the vocalizations of each character.
\item The hash should be communicated in a visual way as well.  This could be done using a hand-written card that is held in the field of view.  If so, the card should be rotated at some point to increase the difficulty of automated tracking and replacement of the card text using video editing software.
\item For maximum security, users may opt to write the hash out during the video onto a pane of glass that is suspended between the user and the recording device, thereby permitting both the users face and the handwriting to remain in view simultaneously at all times.  This would make video editing and replacement of the characters impossible because they are linked to hand motions.  In this case, it is recommended to flip the video horizontally so that the key can be read left to right.
\end{itemize}

\section{Potential attacks} \label{sec_attacks}

In this section we examine the potential attack surfaces and discuss resistances to these attacks.  There are a number of different ways that a man in the middle (MITM) attack might be attempted, so we discuss these methods and the proposed protocol's cryptographic resistances to them separately in the following subsections.

In general, the following MITM attacks would be easiest to perform by the keyserver itself.  However, clients could detect these attempted MITM attacks by periodically querying the keyserver for their own contact address from anonymous IP addresses.  If the wrong public key is returned, then the client would know that a MITM attack is being attempted, and the keyserver would lose credibility.  Thus, even if the technical limitations could somehow be overcome, it is unlikely that a keyserver would be complacent in performing MITM attacks.

If the keyserver is not complacent in MITM attacks, then an attacker would first need to gain access to the user's communication client (i.e., email) in order to complete the key registration procedure with the keyserver before attempting a MITM attack.  Although this might be achievable in special cases by using tactics of social engineering, it is not something that could be easily employed \emph{en masse} for anyone except the communications provider, unless the communications provider itself were compromised or complacent.  In this case, clients would still be able to detect attempted MITM attacks by querying the keyserver and checking the returned public key.

\subsection{Preimage hash attack} \label{attack_preimage}

Theoretically, an attacker could extract the media attestment from a valid identity card and repackage it into a new identity card using a different public key that has the same hash value.  This is known as the \emph{second preimage attack} on the hash function \citep{Rogaway04}.  

For an ideal $n$-bit hash function, the fastest way to compute a second preimage is a brute force enumeration that has time complexity $2^n$, and is generally considered secure for $n \geq 64$.  All currently known practical attacks on MD5 \citep{Stevens07,Vernoux09,Stevens12} are collision attacks.  

Recently, a theoretical attack was published that breaks MD5's preimage resistance, reducing the time complexity of attack from $2^{128}$ to $2^{123.4}$ \citep{Sasaki09, Mao09}.  However, this is still well above the $2^{64}$ needed to be considered secure.  Despite that MD5 is not the most secure hash available, we prefer it to other alternatives for its small output space, which makes reading the hash value in media attestments easier for the user.

\subsection{Media editing attack} \label{attack_media_edit}

Theoretically, an attacker could extract the media attestment from a valid identity card, modify the video to change the communicated hash value (written, spoken, and possibly gestured) to match the hash of the attacker's key, and then repackage it into a new identity card.

The media attestment guidelines (\secref{sec_guidelines}) are specifically designed to make this task sufficiently difficult that users may be confident a media editing attack could not be performed -- at least not with any automated tools.  Indeed, the level of artistic and technical sophistication necessary to convincingly create such an attack is so high that only a professional visual effects company might attempt it, but even then it is dubious whether the results could be convincing.

\subsection{Impostor attack} \label{attack_impostor}

In the impostor attack, the attacker uses a video of himself (or someone else) attesting to the attacker's public key, but claiming the identity and/or contact address (email) of someone else.  This attack would clearly not work against anyone whose identity could be recognized visually, such as a friend, family member, acquintance, or famous person.  Nonetheless, it may be a concern when communicating with online identities.  For this reason, we ask that users show some form of identification (driver's license or passport) in their media attestment.  In addition, a user may examine the rating statistics of other members.

\section{Implementation} \label{sec_implementation}

We have implemented registration and lookup services of the proposed system with a public keyserver that is online at \url{https://www.authma.com}.  Future work will integrate the proposed rating scheme as well as implementing example webmail client using the messaging protocol.

\subsection{Client software support} \label{sec_client_software}

In this section we discuss features that a client software (e.g., a webmail client, mobile phone app, or Facebook app) using the proposed AMA-style key exchange should support.  First, the client's private key should not be stored or retrieved on any third party servers, as this would violate the most fundamental principle of end-to-end encryption.  

If the client is a web client, the key also cannot be stored locally, as this would not be available when changing machines.  Therefore, the most logical way of storing a key is in the user's memory as a passphrase.  A passphrase can be used to derive a public-private keypair at login time by running it through a key expansion algorithm (e.g., PBKDF2 \citep{Kaliski2000}) to generate a seed for a cryptographically secure random number generator that is used to generate an RSA keypair.

Thus, if using a web client, there are necessarily two password fields that must be entered by the user: one to authenticate with the webservice provider, and the other password to be used locally by the client for encryption and decryption.

The client should probably support both unencrypted and encrypted comunication, and only use encryption if both the sender and receiver have generated a keypair.  In order to properly support AMA-style key retrieval, the client should have user-controlled thresholds based on the aggregated community statistics in order to determine if a key can be trusted automatically.  The client must also support functionality for viewing the media attestment, as well as the list of community ratings, means for verifying the signatures on those ratings, and submitting a new rating on a viewed attestment.

After the client establishes trust of a new identity card (which contains public key, media attestment, contact address and user identity), the client should sign this card and cache it for future reference.  This prevents the need for re-querying the keyserver in the future, and signing it prevents the cache entries from being modified by a third party who might have access to the account, such as one's service provider.

In order to support content-based searching of encrypted messages, the client should maintain a search index of all decrypted messages (as well as outgoing messages prior to encryption).  This search index should be encrypted with the user's private key.  If the client is a web client, the search index should be stored in the cloud, so that the client can automatically re-download it when being used from a new machine.

In some cases, a contact address might be used by multiple people, such as a business address like \verb|support@company.com|.  Because this address is used by multiple people, it could not be registered with the keyserver for AMA-style authentication.  However, the owners of this address may still wish to communicate with users via end-to-end encryption.  This can be accomplished if the business encrypts their public key inside the contents of encrypted outgoing messages.  Thus, in order to process such communications seamlessly, a client software should be prepared to extract public keys from incoming messages using an established protocol, as well as embed one's public key in outgoing encrypted messages.

Finally, a client application may wish to query the keyserver for their own address on occassion, from an anonymizing proxy, just to double check that the returned public key is indeed their own public key.  This provides additional security and peace of mind.

\section{Conclusion} \label{sec_conclusion}

Many users would prefer privacy in their communications (as can only be achieved with end-to-end encryption), but not without sacrificing the convenience they have come to expect from webmail clients.  In particular, users don't want the hassle of manually managing public and private keys as typically required for end-to-end encryption.

Although PGP's Web of Trust and open PGP keyservers are designed to mitigate this problem, they cannot be trusted enough for fully automatic key distribution, and hence key management remains a user problem.  Services such as Lavabit and Hushmail have attempted to make PGP more user-friendly by taking care of key management, but in so doing they have given up the fundamental privacy guarantees of end-to-encryption that motivate the use of PGP in the first place.

In this paper, we have proposed a fundamentally new approach by empowering end-users with the capacity to independently verify the authenticity of a public key.  This permits email client software to automatically lookup public keys associated with an email addresses from a keyserver without needing to trust the keyserver, because any MITM attacks could be detected by end-users.  Thus, our protocol enables a new breed of messaging clients with true end-to-end encryption built in, without the hassle of requiring users to manually manage the public keys for their contacts, and without relying on any third party to keep certain information secret.

\bibliographystyle{plainnat}
\bibliography{amakeyx}

\end{multicols}
\end{document}